\documentclass[aps,pra,nofootinbib,preprint,amsmath,amssymb,floatfix]{revtex4}
\usepackage{amssymb}
\usepackage{amsmath}
\usepackage{bm}
\usepackage{upgreek}
\usepackage{graphicx}

\DeclareMathOperator{\Tr}{Tr}

\begin{document}

\title{Presence of negative entropies in Casimir interactions}         

\author{Johan S. H{\o}ye}        

\date{\today}          

\affiliation {Department of Physics, Norwegian University of Science and 
Technology, 7491 Trondheim, Norway,}
\author{Iver Brevik}
\affiliation{Department of Energy and Process Engineering, Norwegian University of
 Science and Technology, 7491 Trondheim, Norway}

\author{Kimball A. Milton}
\affiliation{Homer L. Dodge Department of Physics and Astronomy, Unversity of
Oklahoma, Norman, OK 73019 USA}

\begin{abstract}

Negative entropy in connection with the Casimir effect at uniform temperature 
is a phenomenon rooted in the  circumstance that one is describing a
nonclosed system, or only part of a closed system.
In this paper we show that the phenomenon is not necessarily restricted to 
electromagnetic theory, but can be derived from the quantum theory of 
interacting harmonic oscillators, most typically two oscillators interacting 
not directly but indirectly via a third one. There are two such models, 
actually analogous to the  transverse magnetic (TM) and 
transverse electric (TE) modes in electrodynamics. These mechanical models in 
their simplest version were presented some years ago, by J. S. H{\o}ye et al., 
Physical Review E {\bf 67}, 056116 (2003). In the present  paper we 
re-emphasize the physical significance of the   mechanical picture, and extend 
the theory so as to include the case where there are several mediating 
oscillators, instead of only one.  The TE oscillator exhibits negative
entropy.
Finally, we show explicitly how  the interactions via the electromagnetic field
 contain the two oscillator models.

\end{abstract}
\maketitle

\bigskip
\section{Introduction} \label{sec1}

 Let us begin by recapitulating the conventional procedure for 
 calculating the Casimir force between two dielectric media, typically 
two half-spaces separated by a gap $a$: One starts from the two-point function 
for the electric field at two neighboring spacetime points, usually by using 
the fluctuation-dissipation theorem assuming uniform temperature, and then uses
 Maxwell's stress tensor to calculate the surface pressure, here called $P$. 
Then, the free energy $F$ per unit surface can be found by integration of 
$P=-\partial F/\partial a$, and the internal energy $U$ per unit area follows 
from the thermodynamical formula
\begin{equation}
U=\frac{\partial (\beta F)}{\partial \beta}, \label{intro1}
\end{equation}
with $\beta=1/(k_BT)$. The corresponding entropy $S$ then finally follows from
\begin{equation}
S=\frac{U-F}{T}=  -\frac{\partial F}{\partial T}.
\end{equation}
This procedure is considered in detail at various places, for instance in the 
standard sources  \cite{milton01,bordag09,landau84}.

The following point here calls for  attention: The theory is based upon the  
electrodynamics of a {\it non-closed} physical system. That is, the force is 
calculated from {\tt the} difference in the electrodynamic stress tensor 
between the inside and {\tt the} outside of a dielectric medium. The 
properties of the medium itself are not accounted for. The fact that we are 
dealing with an electromagnetic subsystem makes it not so unreasonable that we 
can encounter unexpected properties when calculating physical properties of the
 subsystem such as  the Casimir entropy.

Consider for definiteness  the two-slab system above,  assuming the separation 
$a$ to be constant. We let the temperature $T$ increase, from zero upwards. 
We further assume the standard Drude dispersion relation
\begin{equation}
\varepsilon (i\zeta)=1+\frac{\omega_p^2 }{\zeta (\zeta +\nu)}, \label{intro2}
\end{equation}
where $\nu$ is the dissipation parameter. As has been shown in detail by 
explicit calculations, as long as $\nu$ is different from zero as always is the
 case for a real material, the slope $\partial F/\partial T$  is {\it zero} at 
$T=0$  \cite{hoye03,brevik06,hoye07}. That is, the Nernst theorem is satisfied 
for the Casimir entropy. We ought to emphasize this point, because assertions 
to the contrary have often appeared in the literature. An ambiguity  might 
occur only if the parameter $\nu$ were exactly zero, {\tt which} is, however, 
only a fictitious case.

Then for increasing temperature the free energy starts to increase while for 
high temperatures it decreases in the usual way.
This increase means that the entropy $S=-\partial F/\partial T$ becomes 
negative in this region. This special property has been subject to several 
studies recently; cf., for instance, Refs.~\cite{lopez12,milton16} with further
 references therein.

 In particular, for high temperature (with the separation $a$ fixed)
the TE contribution is negative, but tends to zero.  This means that the TE
entropy is negative; in fact, it typically is always negative for all
values of $T$.  Whether the total entropy is negative depends on the balance
with the TM entropy, which is typically (but not necessarily)
 positive.  More often that not,
there is a region of low temperature when the total entropy goes negative.

Most previous studies have considered the negative entropy problem from the 
standpoint of electrodynamics. This is quite a natural approach, as the effect 
is related to the circumstance that the relationship between canonical momentum
 $\bf p$ for a particle with mass $m$ and charge $q$ and the electromagnetic 
vector potential ${\bf A(r}, t)$ is $({\bf p}-q{\bf A})^2/2m$ (as is known, 
this is the reason for the absence of classical diamagnetism, the Bohr-van
Leeuwen theorem\footnote{However, recall that the Langevin construction
gives a reasonable model of diamagnetism for dielectrics--see, for example, 
Ref.~\cite{CE}.}).
 It is, however, possible to describe this effect in a different way which is 
simpler and does not involve electromagnetism explicitly, namely as an 
interaction  between two quantum mechanical harmonic oscillators 1 and 2, 
mediated indirectly via a third oscillator 3. Actually we presented this 
oscillator model in an earlier paper (cf. Sec. IV in Ref.~\cite{hoye03}), but 
it seems that this model has been left largely unnoticed. And then we have come
 to the main motivations for the present paper:
\begin{itemize}
\item to re-emphasize the physical significance of the oscillator model;
\item to generalize the theory so as to encompass the case where there are many
 interacting oscillators, similar to the elctromagnetic field, instead of only 
one;
\item to provide a general proof that the TE entropy is negative for high
temperature.
\end{itemize}

\section{Two harmonic oscillator models} \label{sec2}

As mentioned above, we assume the validity of the Drude dispersion relation 
(\ref{intro2}), as this is the most physical one. The competing dispersion 
relation, the plasma relation, corresponds to setting $\nu=0$. The introduction
 of our harmonic oscillator model in Ref.~\cite{hoye03} was actually motivated 
by the current discussion about choosing  between the Drude/plasma 
relations. There are actually two different oscillator models, corresponding to
 the TM and TE modes of the analogous electromagnetic theory.

Consider first the classical partition function of a harmonic oscillator with 
energy
\begin{equation}
H=\frac{1}{2m}p^2+\frac{1}{2}m\omega^2 x^2
\label{7}
\end{equation}
where $x$ is the position, $p$ is the momentum, $\omega$ is the eigenfrequency,
 and $m$ is the mass.  Integrating both momentum and position the classical 
partition function is found to be
\begin{equation}
Z=\frac1{2\pi\hbar}\int\limits_{-\infty}^\infty\int\limits_{-\infty}^\infty 
e^{-\beta H}\,dp\,dx=\frac1{\hbar\beta\omega}
\label{8}
\end{equation}
This gives the free energy and its frequency dependency as
\begin{equation}
F=-\frac{1}{\beta}\ln Z=-\frac{1}{\beta}\ln{\left(\frac1{\hbar\beta\omega}
\right)}\sim \ln \omega.
\label{10}
\end{equation}
Thus for three non-interacting harmonic oscillators the inverse partition 
function is proportional to $\sqrt Q$ where
\begin{equation}
Q=a_1 a_2 a_3, \quad a_i=\omega_i^2\quad(i=1,2,3).
\label{10a}
\end{equation}
By quantization using the path integral method \cite{hoye81,brevik88}, the 
classical system turns out to be split into a set of classical harmonic 
oscillator systems described by Matsubara frequencies. Then for each Matsubara 
frequency expression (\ref{10a}) is replaced by
\begin{equation}
Q=A_1 A_2 A_3, \quad A_i=\omega_i^2+\zeta^2=a_i+\zeta^2,
\label{11}
\end{equation}
where $\zeta=i\omega$. (Depending upon convention $\zeta=-i\omega$ is often 
used.)

Assume now that there is no direct interaction between oscillators 1 and 2. The
 interaction between them is mediated entirely by oscillator 3, which can be 
imagined to be situated in an intermediate position. For simplicity we assume 
all oscillators one-dimensional. The interaction can now be represented as  
$cx_i x_j$ where $x_i$ and $x_j$ are coordinates and $c$ a coupling constant. 
With this the quantity $Q$ becomes
\begin{eqnarray}
Q&=&\left|
\begin{array}{ccc}
A_1 & 0 & c\\
0 & A_2 & c\\
c & c & A_3\\
\end{array}
\right|
=A_1 A_2 A_3 -c^2(A_1+A_2)
\nonumber\\
&=&A_1 A_2 A_3(1-D_1)(1-D_2)\left(1-\frac{D_1 D_2}{(1-D_1)(1-D_2)}\right),
\label{12}
\end{eqnarray}
where
\begin{equation}
D_i=\frac{c^2}{A_i A_3}\quad (i=1,2).
\label{13}
\end{equation}
The quantum free energy $F$ is obtained by summing over the Matsubara 
frequencies $K=\hbar\zeta=i\hbar\omega=2\pi n/\beta$ with $n$ integer
\begin{equation}
\beta F=\lim\frac{1}{2}\sum_n\ln Q(\zeta)\quad(+\mbox{const.}).
\label{14}
\end{equation}
 Here $\lim$ refers to the limit of a discretization procedure.
As pointed out in Ref.~\cite{hoye03}  this must be carefully defined
as in Ref.~\cite{hoye81}
 to obtain correctly the well-known result for $F$. However,  we can skip 
this discussion here as only the last factor of (\ref{12}) is of 
interest. The product $A_1A_2A_3$  represents the three non-interacting 
oscillators. Further the $A_i(1-D_i)$ ($i=1,2$) represent each of the two 
oscillators with their radiation reaction via the third oscillator. Finally the
 last factor represents the induced Casimir energy.

The above model represents the situation analogous to the {\it TM mode}. To
 model the TE mode we will need another model, which is  the analogue to the 
electromagnetic interaction where the third oscillator interacts with the 
{\it momenta}  $p_i$ of the other two, i.e., the interaction $(p_i-\mbox{const}
\times x_3)^2/2m_i$ [$m_i$ is the mass, $i=1,2$]). By evaluation of the 
classical partition function one now will find that the interaction has no 
influence upon thermal equilibrium (as mentioned, this is the analog of 
classical diamagnetism which is equal to zero).  Quantum mechanically the 
problem is less straightforward. But we can simplify  the calculation by 
exchanging the roles of momenta and coordinates of the first two oscillators, 
i.e., we use the  momentum representation. Then the interaction will get the 
form
\begin{equation}
\mbox{const}\times a_i(x_i-\frac{c}{a_i}x_3)^2=\mbox{const}\times(a_i x_i^2-2cx
_i x_3+\frac{c^2}{a_i}x_3^2).
\label{16}
\end{equation}
Compared with the first model considered above an extra $x_3^2$ term has 
appeared with the consequence that the previous coefficient $a_3$ has changed 
to
\begin{equation}
a_3\rightarrow a_3+\frac{c^2}{a_1}+\frac{c^2}{a_2},
\label{17}
\end{equation}
and in the quantum case
\begin{equation}
A_3\rightarrow A_3+\frac{c^2}{a_1}+\frac{c^2}{a_2}.
\label{18}
\end{equation}
Inserted in expression (\ref{12}) this  means that the coefficient $D_i$ has changed to
\begin{equation}
D_i=\frac{c^2}{A_3}\left(\frac{1}{A_i}-\frac{1}{a_i}\right)=
-\frac{\zeta^2c^2}{a_i A_i A_3}.
\label{19}
\end{equation}

Again the free energy due to the interaction follows by summation of the 
logarithm of the last factor of expression (\ref{12}). In the classical high 
temperature limit ($\beta\rightarrow0$) only the term $\hbar\zeta=2\pi 
n/\beta=0$ is present, but with expression (\ref{19}) its contribution is zero.
 This is similar to  what happens for
the TE zero mode ({\tt in the} Drude model) for the Casimir effect. For 
finite temperatures the corresponding
 free energy must be negative. But since it approaches 
zero when $T\rightarrow\infty$, there will be a temperature interval for which 
the Casimir  free energy increases with increasing temperature, 
corresponding  to a Casimir entropy  $S=-\partial F/\partial T$  being  
negative.

\section{Interactions via many oscillators} \label{sec3}

In the models of Sec.~\ref{sec2} two oscillators interacted via a third one. 
This situation we can extend and generalize to interactions via many 
oscillators. Such a situation is the analogue of electromagnetic interactions 
which have a continuum of frequencies. Then the $a_3$ and $A_3$ of 
Eqs.~(\ref{10}) and (\ref{11}) are generalized to
\begin{equation}
a_3\rightarrow a_i,\quad A_3\rightarrow A_i, \quad i=3,4,5,6.\cdots,
\label{20}
\end{equation}
with $a_i=\omega_i^2$ and $A_i=a_i+\zeta^2$ as before.

Again oscillators 1 and 2 interact via oscillators $i$ ($i\geq3$) where the 
coefficient $c$ of Eq.~(\ref{12}) becomes coefficients $c_i$. [Different 
coefficients $c_{1i}$ and $c_{2i}$ for the two oscillators 1 and 2 will also be
possible.] With this one will find that the inverse partition function will be 
the determinant that generalizes Eq.~(\ref{12}) to
\begin{eqnarray}
\nonumber
Q&=&\left|
\begin{array}{ccccc}
A_1 & 0 & c_3 &c_4 & \cdots\\
0 & A_2 & c_3 &c_4 & \cdots\\
c_3& c_3 & A_3 &  0 & \cdots\\
c_4& c_4 & 0 &  A_4 & \cdots\\
\cdots& \cdots& \cdots& \cdots& \cdots\\
\end{array}
\right|
=\left|
\begin{array}{ccccc}
A_1(1-D_1) & -A_1D_1 & c_3 &c_4 & \cdots\\
-A_2D_2& A_2(1-D_2) & c_3 &c_4 & \cdots\\
0 & 0  & A_3 &  0 & \cdots\\
0 & 0 & 0 &  A_4 & \cdots\\
\cdots& \cdots& \cdots& \cdots& \cdots\\
\end{array}
\right|\\
&=&A_1 A_2 (1-D_1)(1-D_2)\left(1-\frac{D_1 D_2}{(1-D_1)(1-D_2)}\right)
\left(\prod\limits_{i\geq3}A_i\right),
\label{21}
\end{eqnarray}
where now
\begin{equation}
D_j=\frac{1}{A_j}\sum_{i\geq3}\frac{c_i^2}{A_i}, \quad(j=1,2).
\label{22}
\end{equation}
Here, to evaluate the determinant, columns $i=3,4,\cdots$ have been multiplied 
with $c_i/A_i$ and subtracted from columns 1 and 2.

The second model is again the analogue of the electromagnetic interaction for 
the TE mode. Then the momenta of oscillators 1 and 2 interact with all the
 oscillators of the electromagnetic interaction. Thus the interaction will have
 the form $(p_j-\sum_{i\geq	3}(c_i x_i))^2/(2m_j)$ ($j=1,2$), and again one
 finds that the interaction has no influence upon the classical partition 
function. To simplify in the quantum case we again can exchange the roles of 
momenta and coordinates of oscillators 1 and 2. Like Eq.~(\ref{16}) the 
interaction then ends up with the form
\begin{equation}
a_j\left(x_j-\frac{1}{a_j}\sum_{i\geq3}\left(c_ix_i\right)\right)^2=a_j x_j^2-2
x_j\sum_{i\geq3}(c_ix_i)+\frac{1}{a_j}
\sum_{i\geq3}\sum_{l\geq3}(c_ic_l x_ix_l)
.\label{23}
\end{equation}
The coefficients $c_i$ can be extended to the more general $c_{ji}$ ($j=1,2$), 
but to simplify the matrices below a bit this is not done. With Eq.~(\ref{23}) 
and short hand notations $\mu=1/a_1+1/a_2$ 
and $q_j=A_j(1/A_j-1/a_j)$, 
Eq.~(\ref{12}) will be generalized to ($A_i=a_i+\zeta^2$)
\begin{eqnarray}
Q&=&\left|
\begin{array}{ccccc}
A_1 & 0 & c_3 &c_4 & \cdots\\
0 & A_2 & c_3 &c_4 & \cdots\\
c_3& c_3 & A_3+c_3^2\mu &  c_3c_4\mu & \cdots\\
c_4& c_4 & c_4c_3\mu &  A_4+c_4c_4\mu & \cdots\\
\cdots& \cdots& \cdots& \cdots& \cdots\\
\end{array}
\right|
=\left|
\begin{array}{ccccc}
A_1 & 0 & c_3 &c_4 & \cdots\\
 0 & A_2 & c_3 &c_4 & \cdots\\
c_3q_1 & c_{\mathbf{3}}q_2  & A_3 &  0 & \cdots\\
c_4q_1 & c_4q_2 & 0 &  A_4 & \cdots\\
\cdots& \cdots& \cdots& \cdots& \cdots\\
\end{array}
\right|.
\label{24}
\end{eqnarray}
In Eq.~(\ref{24}) rows $j=1,2$ have been multiplied with $c_i/a_1$ and 
$c_i/a_2$ respectively and subtracted from rows $i=3,4,\cdots$. Next, 
similar to
Eq.~(\ref{21}) columns $i=3,4,\cdots$ are multiplied with $q_jc_i/A_i$ for 
$j=1,2$ and subtracted from columns 1 and 2. The resulting contributions to the
 inverse partition function is again (\ref{21}), now with $D_j$ given by 
\begin{equation}
D_j=\frac{q_j}{A_j}\sum_{i\geq3}\frac{c_i^2}{A_i}=-\frac{\zeta^2}{a_jA_j}
\sum_{i\geq3}\frac{c_i^2}{A_i}.
\label{25}
\end{equation}
Altogether, this is just a straightforward generalization of result
 (\ref{12}) for $Q$ with $D_j$ either given by Eqs.~(\ref{13}) or (\ref{19}) in
 the two cases. The main difference lies in the quantities $D_j$ that in the 
present section contain many contributions. Thus with $D_j$ given by 
Eq.~(\ref{25}) as with Eq.~(\ref{19}) the corresponding Casimir entropy will be
 negative in an interval as concluded at the end of Sec.~\ref{sec2}.
This constitutes a proof that the TE entropy must always be negative
at high temperature.  Typically, in fact, it is negative at all temperatures
\cite{milton16,ly16}.

An additional notable and interesting feature of the inverse partition 
function, which is the square root of Eq.~(\ref{21})  for each Matsubara 
frequency, is the product
 term for $i\geq 3$. Clearly this part is not affected by the presence of 
oscillators 1 and 2 and their influence upon the resulting eigenfrequencies of 
the coupled system of all oscillators. Thus oscillators $i\geq3$ can without 
any approximation be eliminated or disregarded to be replaced by the 
interaction quantities $D_j$ at thermal equilibrium. Correspondingly,
 with polarizable 
media the quantized electromagnetic field can be eliminated to be replaced by 
the radiating dipole-dipole interaction. This simplification we have utilized 
in Ref.~\cite{brevik88} and later works.

\section{Interaction via the electromagnetic dipole radiation field} 
\label{sec4}

For two oscillators interacting via the electromagnetic field it should now be 
possible to identify this situation with Eq.~(\ref{21}) where $D_j$ is 
 expression (\ref{22}) for the TM mode  and expression (\ref{25}) for the
TE mode. The free energy of 
interaction (Casimir energy) follows from the logarithm  of the
penultimate factor of 
(\ref{21}) when inserted in Eq.~(\ref{14}). As we will see, the radiating 
dipole interaction has the form and structure consistent with the expressions 
for $D_j$.

For two oscillators interacting via the potential $\psi({\bf r}) s_1 s_2$ with 
oscillator coordinates  $s_i$, which can be identified with 
polarization (here in one dimension
 for simplicity). The  Casimir free energy $F$ in the classical case is given 
by Eq.~(3.4) in Ref.~\cite{brevik88} as
\begin{equation}
\beta F=\frac{1}{2}\ln(1-(\alpha\psi)^2)\approx-\frac{1}{2}
(\alpha\psi({\bf r}))^2
\label{30}
\end{equation}
where $\alpha$ is polarizability. In the quantum case one sums over Matsubara 
frequencies as in  Eq.~(5.8) of that reference by which
\begin{equation}
\beta F=\frac{1}{2}\sum _K\ln(1-(\alpha_K\psi_K)^2).
\label{31}
\end{equation}

With two equal oscillators (same $\alpha$) it should be possible to make the 
identification
\begin{equation}
\frac{D_j}{1-D_j}\rightarrow\alpha_K\psi_K.
\label{32}
\end{equation}
It is clear that $1/A_j$ corresponds to $\alpha_K\propto1(a_j+\zeta^2)$ for a 
simple oscillator with eigenfrequency $\omega=\sqrt{a_j}$ not interacting with 
its surroundings. As pointed to below Eq.~(\ref{12}) the $A_j(1-D_j)$ represent
s oscillator $j=1$ or 2 alone and their interactions with oscillator 3. Thus in
 Eq. (\ref{21}) the same factor represent the interaction of oscillator $j$ 
with the electromagnetic field represented by oscillators $i\geq 3$. So 
$1/(A_j(1-D_j))$ corresponds to $\alpha_K$ with radiation reaction taken into 
account. With this the remaining part $A_j D_j$ of (\ref{31}) should represent 
$\psi_K$. According to Eqs.~(\ref{22}) or (\ref{25}) this gets contributions 
from the oscillators through which oscillators 1 and 2 interact. Then the 
remaining crucial question is whether the radiating dipole interaction $\psi_K$
 is consistent with the two expressions for $D_j$. Thus we must look for the 
eigenmodes of the electromagnetic field. In free space these modes are plane 
waves of wave vector ${\bf k}$ and frequency
\begin{equation}
\omega=ck
\label{33}
\end{equation}
where $c$ is light velocity. These waves should, if possible, be identified 
with the oscillators $i\geq0$ of Sec.~\ref{sec3}. And this identification we 
find from the Fourier transform of the radiating dipole interaction. This 
interaction is given by Eq.~(6.1) in Ref.~\cite{hoye98} ($\zeta=i\omega$)
\begin{equation}
\tilde \phi(12)=\frac{4\pi}{3}s_1 s_2\frac{1}{(ck)^2+\zeta^2}[(ck)^2 \tilde 
D(12)+2\zeta^2\, \hat {\bf s}_1\cdot\hat {\bf s}_2]\quad(+\rm{const.}),
\label{34}
\end{equation}
with $\tilde D(12)=3(\hat{\bf k}\cdot \hat{\bf s}_1)(\hat{\bf k}\cdot 
\hat{\bf s}_2)-\hat {\bf s}_1\cdot\hat {\bf s}_2$. The hats denote unit 
vectors. Here ${\bf s}_j$ are the polarizations of the two oscillators. The 
constant term can be disregarded as it only contributes to a 
$\delta$-function $\delta (r)$ in ${\bf r}$-space and is thus zero anyway with 
$r\neq0$.

It is now easily seen that expression (\ref{34}) has precisely the form where 
both expressions (\ref{22}) and (\ref{25}) for $A_j D_j$  are present with 
$A_i$ given by (\ref{11}). The $\tilde D(12)$ and $\hat {\bf s}_1\cdot\hat 
{\bf s}_2$ terms of expression (\ref{34}) correspond to expressions (\ref{22}) 
and (\ref{25}) respectively. With Fourier transform (\ref{34}) the frequency 
dependent dipole interaction $\psi(k)\rightarrow\phi(12)$ is given by
\begin{equation}
\phi(12)=\frac{1}{(2\pi)^3}\int\tilde\phi(12)e^{i{\bf kr}}\,d{\bf k}=
s_1s_2\left[\psi_{DK}
(r) D(12)+\psi_{\Delta K}(r)\Delta(12)\right],
\label{35}
\end{equation}
where from Eq.~(5.10) of Ref.~\cite{brevik88}
\begin{equation}
\psi_{DK}(r)=-\frac{e^{-\tau}}{r^3}\left(1+\tau+\frac{1}{3}\tau^2\right) 
\quad\mbox{and}\quad \psi_{\Delta K}(r)=-\frac{2e^{-\tau}}{3r^3}\tau^2 \quad 
(+\mbox{const.}\,\delta({\bf r})),
\label{36}
\end{equation}
with $\tau=i\omega r/c$.

Thus altogether, interactions via the electromagnetic field contain both the 
two oscillator models considered in Sec.~\ref{sec3}. The dipole-dipole 
interaction (\ref{34}) is then a sum ($\rightarrow$ integral) of eigenmodes 
(Fourier components) that induce the resulting interaction between the two 
oscillators. An implication of this, as we have seen, is  that the contribution
 to the entropy can be negative in some regions.

\section{Summary}

We have studied the reason for possible negative entropy related to the Casimir
 interaction between two media. This negative entropy may seem unphysical. To 
show that this is not so, we have studied two harmonic oscillator models where 
two oscillators interact via a third one. For one of the models the momenta of 
the two oscillators interact with the amplitude of the third one in a way 
similar to the interaction with the electromagnetic vector potential, and
in fact corresponds to the TE polarization.  Then a 
negative entropy contribution is found. This shows that this type of behavior 
is not unphysical. Then the situation with the third oscillator is generalized 
to a set of oscillators that mediates the induced interaction between the two 
oscillators. Finally it is noticed that the induced radiating dipole-dipole 
interaction between a pair of oscillating dipole moments can be identified 
with a combination of the induced ones of the two oscillator models.
This paper gives a proof that the TE contribution to the entropy
must be negative for large $T$, being typically negative for all $T$.  The
TM contribution is typically positive.  The total entropy, therefore, is likely
to contain a negative entropy region.

\appendix
\section{Field theory approach}
The point of this appendix is to show that the considerations of the main text
have a close correspondence with the  field theoretic approach in quantum
electrodynamics.  The latter starts from the expression for the free energy
as a sum over Matsubara frequencies (here $\hbar=c=1$)
\begin{equation}
F=-\frac{T}2\sum_{n=-\infty}^\infty \Tr \ln \bm{\Gamma \Gamma}_0^{-1},
\end{equation}
where $\bm{\Gamma}_0$ is the free electromagnetic Green's dyadic, and 
$\bm{\Gamma}$ is that in the presence of bodies which interact with the
electromagnetic field, e.g., dielectric or metallic bodies.  For the case
of dielectrics, we can define a potential in terms of the permittivity 
$\varepsilon$, $V=\varepsilon-1$, and then we can readily show for two
disjoint bodies, for which $V=V_1+V_2$, that the free energy is
\begin{equation}
F=\frac{T}2\sum_n\Tr\ln[(\bm{1}-\bm{\Gamma}_0V_1)(\bm{1}-\bm{\Gamma}_0
\mathbf{T}_1\bm{\Gamma}_0\mathbf{T_2})(\bm{1}-\bm{\Gamma}_0V_2)],\label{tgtg}
\end{equation}
in terms of the scattering matrices
\begin{equation}
\mathbf{T}_i=V_i(\bm{1}-\bm{\Gamma}_0V_i)^{-1}.
\end{equation}
Evidently, Eq.~(\ref{tgtg}), sometimes called the TGTG formula, is identical
with Eq.~(\ref{21}) inserted into Eq.~(\ref{14}), which was derived long before
the modern renaissance of multiple-scattering formulations of Casimir
problems.  Here the $A_i$'s have been disregarded, as not involving interaction
with the electromagnetic field, and the $D_i$ are identified with
\begin{equation}
D_i\leftrightarrow \bm{\Gamma}_0 V_i.
\end{equation}

And the break-up into electromagnetic modes, detailed in Sec.~\ref{sec4}, is
just the well-known decomposition
\begin{eqnarray}
\bm{\Gamma}_0(\mathbf{r})
&=&(\bm{\nabla\nabla-1}\nabla^2)\frac{e^{-|\zeta_n|r}}{r}
\nonumber\\
&=&\left[(3\mathbf{\hat r\hat r}-\bm{1})\left(1+|\zeta|r+\frac13\zeta^2r^2
\right)-\bm{1}\frac23\zeta^2r^2\right]\frac{e^{-|\zeta|r}}{r^3},
\end{eqnarray}
which restates Eqs.~(\ref{35}) and (\ref{36}).
So the correspondence is not merely analogous, it is precise.

\end{document}